\documentclass{article}
\usepackage{graphicx}
\usepackage{here}

\usepackage{comment}
\usepackage{bm}
\usepackage{ascmac,amssymb,amsfonts, amsmath, amsthm}

\usepackage{algorithm, algpseudocode} 

\renewcommand{\algorithmicrequire}{\textbf{Input:}}
\renewcommand{\algorithmicensure}{\textbf{Output:}}
\algnewcommand{\Initialize}[1]{%
	\State \textbf{Initialize:}
	\Statex \hspace*{\algorithmicindent}\parbox[t]{0.8\linewidth}{\raggedright #1}
}
\algnewcommand{\RETURN}[1]{\State \algorithmicreturn~ #1}
\usepackage{etoolbox}
\newcommand*{\affaddr}[1]{#1} 
\newcommand*{\affmark}[1][*]{\textsuperscript{#1}}

\usepackage{natbib}

\usepackage{mathrsfs}

\newcommand{\argmin}{\mathop{\rm arg~min}\limits}

\newcommand{\xr}{\bm r}

\newcommand{\xC}{\bm C}
\newcommand{\xD}{\bm D}

\newcommand{\xb}{\bm b}
\newcommand{\xd}{\bm d}

\newcommand{\xsigma}{\bm \sigma}
\newcommand{\xu}{\bm u}
\newcommand{\xv}{\bm v}
\newcommand{\xw}{\bm w}

\newcommand{\xo}{\bm o}
\newcommand{\xp}{\bm p}

\newcommand{\xzero}{\bm 0}
\newcommand{\xone}{\bm 1}
\newcommand{\xa}{\bm a}

\newcommand{\mj}{\mathcal{J}}

\newcommand{\trans}{\mathsf{T}}

\newtheorem{thm}{Theorem}

\newtheorem{lem}{Lemma}

\newtheorem{asum}{Assumption}

\newtheorem{defi}{Definition}

\newenvironment{myproof}{\paragraph{\bf Proof.}}{\hfill$\qed$}
\newenvironment{myproofname}[2] {\paragraph{\bf Proof of {#1} {#2}.}}{\hfill$\qed$}

\begin{document}
\title{A 3-approximation list scheduling algorithm for a single-machine scheduling problem with a non-renewable resource and total weighted completion time criterion
\thanks{This research is supported in part by Grant-in-Aid for Science
	Research (A) 19H00808 of Japan Society for the Promotion of Science.}}


\author{Susumu Hashimoto
	\thanks{Department of Industrial Engineering and Economics, Tokyo Institute of Technology, Meguro-ku, Tokyo, Japan. Email:hashimoto.s.ak@m.titech.ac.jp}
\and
Shinji Mizuno
	\thanks{Department of Industrial Engineering and Economics, Tokyo Institute of Technology, Meguro-ku, Tokyo, Japan. Email:\texttt{mizuno.s.ab@m.titech.ac.jp}}}

%
%
\date{May 3, 2021}

\maketitle

\begin{abstract}

This paper studies a single-machine scheduling problem with a non-renewable resource (NR-SSP) and total weighted completion time criterion. The non-renewable resource is consumed when the machine starts processing a job. We consider the case where each job's weight in the objective function is proportional to its resource consumption amount. The problem is known to be NP-hard in this case. We propose a 3-approximation list scheduling algorithm for this problem. Besides, we show that the approximation ratio 3 is tight for the algorithm.

\end{abstract}

\section{Introduction}
This paper proposes a 3-approximation list scheduling algorithm for a single-machine scheduling problem with a non-renewable resource (NR-SSP). The non-renewable resource is consumed when the machine starts processing a job. The resource is provided at fixed times and in fixed amounts. We consider the case where each job's weight in the objective function is proportional to its resource consumption amount. Such a condition appears in some production systems with a single product. This problem is more general than NR-SSPs studied by \cite{gyorgyi2019minimizing}, \cite{gyorgyi2020new}, and \cite{berczi2020scheduling}. Besides, no constant-factor approximation algorithm is known for the problem to our knowledge.

NR-SSP is the problem of deciding start (or completion) times of $ n $ jobs on a machine under given $ q $ times resource supplies. We denote the jobs by $ \mj = (J_1, J_2, \dots, J_n) $. Each job $ J_j $ has a processing time $ p_j > 0 $, a weight $ w_j>0 $, and a resource requirement $ a_j > 0 $. Each provision plan has a supply time $ u_i \ge 0 $ and a supply amount $ b_i > 0 $. The machine cannot process multiple jobs simultaneously, and it cannot preempt any process. Besides, each job $ J_j $ consumes $ a_j $ resources to start processing. The machine cannot start processing a job if the resources are not enough. This condition is expressed as follows:
\begin{equation}
	\label{eq:balance}
	\sum_{j:C_j - p_j \le T}a_j \le \sum_{i:u_i \le T}b_i \ \mbox{for any} \  T \ge 0,
\end{equation}
where $ C_1,C_2, \dots, C_n $ indicate completion times of jobs $ J_1, J_2, \dots, J_n $.
The objective function is total weighted completion time, which is expressed as $\sum_{j=1}^n{w_j C_j}$.
We introduce the triplet notation $ \alpha|\beta|\gamma $ for scheduling problems by \cite{graham1979optimization}, where $ \alpha $ stands for machine environments, $ \beta $ represents other conditions, and $ \gamma $ is the objective function. The problem mentioned above is expressed as $ 1|nr=1, a_j=w_j|\sum w_j C_j $ by using this notation, where $ \alpha=1 $ means 1-machine problem, $ nr $ represents the number of sorts of non-renewable resources, $ a_j=w_j $ indicates $ a_j=w_j $ for all $ j \in \{1, 2, \dots, n\} $, and $ \sum w_j C_j$ stands for that the objective functon is $\sum_{j=1}^n w_j C_j$.

Next, we summarize known research results for special cases of $ 1|nr=1, a_j=w_j|\sum w_j C_j $.
\cite{berczi2020scheduling} prove that $ 1|nr=1, a_j=1|\sum C_j $ is strongly NP-hard problem.
Hence $ 1|nr=1, a_j=w_j|\sum w_j C_j $ is also strongly NP-hard problem.
Although there are no explicit theoretical results about approximation algorithms for this problem, results for more limited problems are known.
\cite{berczi2020scheduling} propose $ \frac{3}{2} $-approximation algorithm for $ 1|nr=1, a_j=1|\sum C_j $.
\cite{gyorgyi2019minimizing} prove that $ 1|nr=1, p_j=1, w_j=a_j, q=2|\sum w_jC_j $ is weakly NP-hard problem.
Moreover, they give a 2-approximation list scheduling algorithm for $ 1|nr=1, w_j=p_j=a_j|\sum w_jC_j $.
\cite{gyorgyi2020new} propose a 2-approximation list scheduling algorithm for $ 1|nr=1, p_j=1, w_j=a_j, q=2|\sum w_jC_j $, and they show that the algorithm is a 3-approximation algorithm for $1|nr=1, p_j=1, w_j=a_j|\sum w_jC_j $.
They also conjecture that the approximation ratio can be reduced to 2.
\cite{hashimoto2021tight} prove that the list scheduling algorithm by \cite{gyorgyi2020new} is actually a 2-approximation algorithm for $ 1|nr=1, p_j=1, w_j=a_j|\sum w_jC_j $.

For other NR-SSP problems, we mention some complexity and (or) approximation results.
\cite{grigoriev2005basic} show that $ 1|nr=k|C_{\max} $ is strongly NP-hard for any integer $ k $, where $ C_{\max} $ represents the maximum completion time or makespan. Besides, they prove that any list scheduling algorithm is a 2-approximation algorithm for the problem.
\cite{gafarov2011single} prove that $ 1|nr=1|\sum C_j $, $ 1|nr=1|\sum U_j $, and $ 1|nr=1|L_{\max} $ are strongly NP-hard, where $ \sum U_j $ stands for the number of late jobs and $L_{\max}$ indicates the maximum lateness.
\cite{gyorgyi2015approximability} propose PTASs for $ 1|nr=const., q=const.|C_{\max} $ and $ 1|nr=1, a_j=p_j, r_j|C_{\max} $, where $ r_j $ means that the problem has release dates.
\cite{kis2015approximability} proves that $ 1|nr=1, q=2|\sum w_jC_j $ is NP-hard, and establishes an FPTAS for the problem.
\cite{berczi2020scheduling} establish a 6-approximation algorithm and a PTAS for $ 1|nr=1, p_j=0|\sum w_jC_j $.

In addition to NR-SSP, scheduling problems with non-renewable resource(s) have also been studied.
\cite{carlier1982scheduling} develop an exact polynomial-time algorithm for a precedence constrained scheduling problem with a non-renewable resource. \cite{slowinski1984preemptive} gives an exact algorithm for a preemptive parallel machine scheduling problem with a non-renewable resource.
\cite{briskorn2013exact} propose exact algorithms for a single-machine inventory constrained scheduling problem with total weighted completion time criterion. \cite{morsy2015approximation} give 2-approximation algorithms for limited cases of the above inventory constrained scheduling problem.

\section{Algorithms for NR-SSP}
We denote any instance of a single-machine scheduling problem with a non-renewable resource (NR-SSP) by $S=\langle n,\mj, \xp, \xw, \xa, q, \xu, \xb \rangle$,
where $n$ is the number of jobs,
$\mj=( J_1, J_2, \dots, J_n ) $ is a sequence of jobs,
$\xp = (p_1,p_2,\dots,p_n)^\trans $ is a (column) vector of processing times,
$\xw = (w_1,w_2,\dots,w_n)^\trans$ is a vector of weights,
$\xa = (a_1,a_2,\dots,a_n)^\trans$ is a vector of resource requirements,
$q$ is the number of supply times,
$\xb = (b_1,b_2,\dots,b_q)^\trans$ is a vector of supply amounts,
and $\xu = (u_1,u_2,\dots,u_q)^\trans$ is a vector of supply times, where $ (\cdot)^\trans $ denotes transpose.
In this paper, we deal with the case where $\xw=\xa$.
Then the instance is simply denoted by $S=\langle n,\mj, \xp, \xa, q, \xu, \xb \rangle$.
A schedule for $S$ is denoted by $\xC=(C_1, C_2, \ldots, C_n)^\trans$ which is a vector of completion times of the jobs.
The objective function of the schedule is
\[
 f_S(\xC) = \sum_{j=1}^n w_j C_j = \xw^\trans \xC = \xa^\trans \xC.
\]
The problem NR-SSP is to find a feasible schedule which minimize this objective function.
Throughout the paper, we assume that there are no resources other than $\xb$,
\[
 n \ge 1, \xp > \xzero_n, \xw=\xa > \xzero_n, q \ge 1, \xb > \xzero_q, u_1 \ge 0,
\]
and
\[
 u_{j+1} > u_{j} \quad \mbox{for any} \quad j \in \{1,2,\dots,q-1 \},
\]
where $\xzero_n$ denotes the $n$-dimensional zero vector.

Let $ N = \{1,2, \dots, n\} $ and $ \xr =(r_1, r_2, \dots, r_n)^\trans$ be the vector of $r_i= a_i/ p_i$ for $i \in N$.
We denote an order of jobs $ J_i, i \in N$ by using a permutation $\xo$ of $ (1,2, \dots, n) $.
Let $ {\mathcal O} $ be the set of all the permutations of $ (1,2, \dots, n) $.
Then we define a subset of $ {\mathcal O} $ for two vectors $\xa$ and $\xp$ in any instance $S=\langle n,\mj, \xp, \xa, q, \xu, \xb \rangle$ as follows.
\begin{defi}
\label{def.o(a,p)}
For $\xa$ and $\xp$ in any instance $S=\langle n,\mj, \xp, \xa, q, \xu, \xb \rangle$,
we define a set $O(\xa, \xp) \subset {\mathcal O} $ of permutations $\xo$ satisfing the following conditions:
\begin{itemize}
\item[(i)]
 $a_{o(n)} = \min_{k \in N} a_k$,
\item[(ii)]
	for any $ j \in \{1,2, \dots, n-1\} $, if $a_{o(j)} > \sum_{k=j+1}^na_{o(k)}$, then
	\[
	 a_{o(i)} \ge a_{o(j)} \ \mbox{for any} \ i \in \{1,2, \dots, j \}, 
	\]
\item[(iii)]
	for any $ j \in\{1,2, \dots, n-1\} $, if $a_{o(j)} \le \sum_{k=j+1}^na_{o(k)} $, then
	\begin{alignat*}{2}
	 r_{o(i)} \ge r_{o(j)} \ &\mbox{for any} \ i \in \{1,2, \dots, j \}\\ &\mbox{such that} \ a_{o(i)} \le \sum_{k=j+1}^na_{o(k)}.
	\end{alignat*}
\end{itemize}
\end{defi}

Now we propose two algorithms, Algorithm \ref{alg:order} and Algorithm \ref{alg:list}.
Algorithm \ref{alg:order} outputs a permutation $\xo \in O(\xa,\xp)$ for inputs $\xa, \xp$ in $S=\langle n,\mj, \xp, \xa, q, \xu, \xb \rangle$.
Algorithm \ref{alg:list} is a simple list scheduling algorithm which outputs a vector $\xC$ of completion times for inputs $S=\langle n,\mj, \xp, \xa, q, \xu, \xb \rangle$ and $\xo \in {\mathcal O}$.

Algorithm \ref{alg:order} is motivated by the 6-approximation algorithm in \cite{berczi2020scheduling}. Their algorithm determines a job's processing order in inverse. It chooses the least economical ($ \argmin w_j/a_j $) job at each selection except for jobs whose weight is bigger than the total weight of selected jobs.
In our algorithm, we employ $ \argmin w_j/p_j $ instead of $ \argmin w_j/a_j $ as the economic indicator.

\begin{algorithm}[H]
	\caption{A job-ordering algorithm for NR-SSP}
	\label{alg:order}
	\begin{algorithmic} [1]
	\Require {$\xa, \xp$ in any instance $S=\langle n,\mj, \xp, \xa, q, \xu, \xb \rangle$ of NR-SSP} 
	\Ensure{ $\xo=(o(1),o(2),\dots, o(n)) \in {\mathcal O}$ }
	\Statex
	\Initialize{\Statex $SUMW \gets 0$\\
	$ N^{REST} \gets N $}
	\For{$ i = n $ to $ 1 $} \Comment{Inverse order}
	\State{$ N^\prime \gets \left\{j\in N^{REST}|a_j \le SUMW \right\} $}\label{step:n_rest}
	\If{$ N^\prime = \emptyset $}
	\State{$ o(i) \gets \argmin_{j \in N^{REST}}a_j $( ties are broken arbitrarily)}\label{step:big_weight}
	\Else
	\State{$ o(i) \gets \argmin_{j \in N^\prime}r_j $( ties are broken arbitrarily)}\label{step:small_weight}
	\EndIf
	\State{$ SUMW \gets SUMW + a_{o(i)} $}
	\State{$ N^{REST} \gets N^{REST} \setminus \{o(i)\} $}
	\EndFor
	\end{algorithmic}
\end{algorithm}
\begin{algorithm}[H]
	\caption{A list-scheduling algorithm for NR-SSP}
	\label{alg:list}
	\begin{algorithmic}[1]
		\Require{$S=\langle n,\mj, \xp, \xa, q, \xu, \xb \rangle$ and $\xo=(o(1),o(2),\dots, o(n)) \in {\mathcal O}$}
		\Ensure$\xC =(C_1,C_2,\dots,C_n)^\trans$ \Comment{a schedule}
		\Statex
		\If{$ \sum_{j=1}^{n}a_j>\sum_{q=1}^{m}b_q $}
		\RETURN{INFEASIBLE} \Comment{The instance is infeasible}
		\EndIf
		\State{$ C_{o(1)} \gets \min\{ T | a_{o(1)} \le \sum_{i:u_i \le T} b_i \} + p_{o(1)}$}
		\For {$j = 2$ to $n$}
		\State{\[
			C_{o(j)} \gets \min \left\{ T \middle| T \ge C_{o(j-1)},\sum_{i=1}^{j} a_{o(i)} \le \sum_{i:u_i \le T} b_i \right \} + p_{o(j)}.
			\]}
		\EndFor
	\end{algorithmic}
\end{algorithm}
For any instance $S=\langle n,\mj, \xp, \xa, q, \xu, \xb \rangle$,
we denote an optimal schedule by $\xC^*(S) = (C_1^*(S),C_2^*(S), \dots, C_n^*(S))^\trans$
and the output schedule of Algorithm \ref{alg:list} for any $ \xo \in {\mathcal O} $ by $\xC^{\xo}(S) = (C_1^{\xo}(S),C_2^{\xo}(S), \dots, C_n^{\xo}(S))^\trans$.

\section{Main results for Algorithms 1 and 2}

In the rest of the paper,
we impose the following assumptions for any instance $ S=\langle n, \mj, \xp, \xa, q, \xu, \xb\rangle $ without loss of generality:
\begin{asum}
	\label{asum:sum}
	$\sum_{j=1}^n a_j = \sum_{i=1}^q b_i$. 
\end{asum}

\begin{asum}
	\label{asum:min_ratio}
	$ \max_{j \in N} \frac{a_j}{p_j} \le1 $.
\end{asum}

\begin{asum}
	\label{asum:indopt}
	Jobs are sorted such that $ C_1^*(S)< C_2^*(S)<\cdots<C_n^*(S)$.
\end{asum}
Since we do not know the optimal schedule $ \xC^*(S)$, we use Assumption \ref{asum:indopt} only for simplicity of discussion.

For any instance $S=\langle n,\mj, \xp, \xa, q, \xu, \xb \rangle$ and any $\xo \in O(\xa,\xp)$,
we obtain the following results for the output schedule $\xC^{\xo}(S)$ of Algorithm 2.
The results are proved in the succeeding sections.
\begin{thm}
	\label{thm:main}
For any instance $S=\langle n,\mj, \xp, \xa, q, \xu, \xb \rangle$ of NR-SSP and any permutation $ \xo \in O(\xa, \xp) $,
 $ f_S(\xC^{\xo}(S))<3f_S(\xC^*(S)) $ holds.
\end{thm}
\begin{thm}
 \label{thm:lowerbound}
For any $\epsilon>0$, there exists an instance $\tilde S=\langle n,\mj, \xp, \xa, q, \xu, \xb \rangle$
of NR-SSP and an $ \xo \in O(\xa,\xp) $ such that $ f_S(\xC^{\xo}(\tilde S)) \ge (3-\epsilon) f_S(\xC^*(\tilde S)) $.
\end{thm}
Theorem \ref{thm:main} states that a combination of Algorithms 1 and 2 is a 3-approximation algorithm and
Theorem \ref{thm:lowerbound} states that the approximation ratio 3 is tight.

The next three theorems are used to prove Theorem \ref{thm:main}.
\begin{thm}
	\label{thm:just_in_time}
Let $S=\langle n,\mj, \xp, \xa, q, \xu, \xb \rangle$ be any instance of NR-SSP. Define $ S^\prime = \langle n,\mj, \xp, \xa, n, (\xC^*(S) - \xp), \xa\rangle$.
For any $ \xo \in {\mathcal O} $, $ f_{S^\prime}(\xC^{\xo}(S^\prime)) < 3f_{S^\prime}(\xC^{*}(S^\prime))$ implies $ f_{S}(\xC^{\xo}(S)) < 3f_{S}(\xC^{*}(S)) $.
\end{thm}
\begin{thm}
	\label{thm:translation}
Let $ S = \langle n, \mj, \xp, \xa, n, (\xC^\prime - \xp), \xa\rangle $ be any instance of NR-SSP,
where $ \xC^\prime=(C_1^\prime, C_2^\prime, \dots, C_n^\prime)^\trans $ 
such that $ C_1^\prime \ge p_1 $
and $ C_{j+1}^\prime-C_j^\prime\ge p_{j+1} $ for any $ j \in \{1,2,\dots, n-1\} $.
Let $ \xsigma(\xp) = (p_1, p_1+p_2, \dots, \sum_{j = 1}^n p_j)^\trans $ and $ \bar{S} = \langle n, \mj, \xp, \xa, n, (\xsigma(\xp) - \xp), \xa\rangle$.
For any $ \xo \in O(\xa,\xp) $, $ f_{\bar{S}}(\xC^{\xo}(\bar{S})) < 3f_{\bar{S}}(\xC^{*}(\bar{S})) $ implies $ f_{S}(\xC^{\xo}(S)) < 3f_{S}(\xC^{*}(S)) $.
\end{thm}
\begin{thm}
	\label{thm:same_ratio}
	Let $ S = \langle n, \mj, \xp, \xa, n, (\xsigma(\xp) - \xp), \xa\rangle$ where $ \xsigma(\xp) = (p_1, p_1+p_2, \dots, \sum_{j = 1}^n p_j)^\trans $. Define $ S^\dagger = \langle n, \mj, \xa, \xa, n, (\xsigma(\xa) - \xa), \xa\rangle$.
Then $ O(\xa, \xp) \subset O(\xa,\xa) $ and,
for any $ \xo \in O(\xa, \xp) $, $ f_{S^\dagger}(\xC^{\xo}(S^\dagger)) < 3f_{S^\dagger}(\xC^{*}(S^\dagger))$ implies $f_{S}(\xC^{\xo}(S)) < 3f_{S}(\xC^{*}(S)) $.
\end{thm}
Theorem \ref{thm:just_in_time} states that it suffices to prove Theorem \ref{thm:main} only in the case of $\langle n,\mj, \xp, \xa, n, (\xC^*(S) - \xp), \xa\rangle$. Besides, Theorem \ref{thm:translation} and \ref{thm:same_ratio} ensures that we may fix $\xu = \xsigma(\xp) - \xp$ and $\xa =\xp$. From these theorems, it is enough to prove Theorem \ref{thm:main} for the special instance $S = \langle n,\mj, \xa, \xa, n, (\xsigma(\xa) - \xa), \xa \rangle$.

The next two theorems show some properties of any permutation $\xo \in O(\xa,\xp)$ and
we use them in the proof of Theorems \ref{thm:translation} and \ref{thm:same_ratio}.
We introduce some notations to state the theorems. 
For any instance $S=\langle n,\mj, \xp, \xa, q, \xu, \xb \rangle$ and any permutation $ \xo \in {\mathcal O} $,
we define
\[
 A^*_j= \sum_{k=j}^na_k \ \mbox{and} \ A^{\xo}_j=\sum_{k=j}^na_{o(k)} \ \mbox{for} \  j\in N.
 \]
For simplicity of discussion, let $ A^*_{j}=A^{\xo}_{j}=0 $ for any $ j \ge n+1 $.
Then we define
\[
 \lambda_j =\min \{ {k\in N} | A^{\xo}_k \ge A^*_j \} \ \mbox{for any} \ j\in \{1,2, \dots, n+1\}.
 \]
From the definition, we see that
\[ 
  A^{\xo}_{{\lambda_j}+1} < A^*_j \le A^{\xo}_{\lambda_j} \ \mbox{for any } \ j \in N.
\] 

\begin{thm}
\label{prop:2opt_condition}
	Let $ S = \langle n,\mj, \xp, \xa, q, \xu, \xb \rangle$ be any instance of NR-SSP and $ \xo \in O(\xa, \xp)$,
	then $ A^{\xo}_{\lambda_j} \le 2A^*_{j}$ for any $ j \in \{1,2, \dots, n+1\} $.
\end{thm}

\begin{thm}
	\label{prop:delta_inequality}
	Let $ S = \langle n,\mj, \xp, \xa, q, \xu, \xb \rangle$ be any instance of NR-SSP and $\xo \in O(\xa,\xp)$.
	For any $k, l \in N$,
\begin{itemize}
\item[(i)] if $ r_{o(k)} \ge r_{o(l)} $, then
$a_{o(l)} ( p_{o(k)}-a_{o(k)}) \le a_{o(k)} ( p_{o(l)}-a_{o(l)}) $,
\item[(ii)] if $k< l$ and $ r_{o(k)} < r_{o(l)} $, then
 $a_{o(k)} \ge a_{o(l)} \ \mbox{and} \ a_{o(k)} \ge A^{\xo}_{l+1}$.
\end{itemize}
\end{thm}

\section{Proof of Theorems \ref{prop:2opt_condition} and \ref{prop:delta_inequality} }

In this section, we prove Theorems \ref{prop:2opt_condition} and \ref{prop:delta_inequality}.
\begin{myproofname}{Theorem}{\ref{prop:2opt_condition}}
If $ j=n+1 $, then $A^{\xo}_{\lambda_{n+1}} = 2A^*_{n+1} = 0$. Thus we assume that $ j \in N $.
	From the definition of $\lambda_j$,
	\begin{equation}
		\label{eq:2opt_1}
		A^{\xo}_{\lambda_j+1} < A^*_{j} = \sum_{k=j}^n a_k.
		\end{equation}
	If $a_{o(\lambda_j)}\le A^{\xo}_{\lambda_j+1}$, then $A^{\xo}_{\lambda_j} = a_{o(\lambda_j)} + A^{\xo}_{\lambda_j+1} < 2 A^*_j$; 
	therefore we assume $a_{o(\lambda_j)}> A^{\xo}_{\lambda_j+1}$.
	From Definition \ref{def.o(a,p)}, 
	 we see $ a_{o(\lambda_j)} \le a_{o(k)}$ for any $k \in \{1, 2, \dots, \lambda_j \} $.
	The inequality \eqref{eq:2opt_1} implies that $ A^{\xo}_{\lambda_j+1} $ does not have at least one term $a_{o(k)}$ in $ A^*_{j} $.
	Hence we see $a_{o(\lambda_j)} \le a_{o(k)} \le A^*_{j}$. From this inequality and \eqref{eq:2opt_1}, we obtain
	$A^{\xo}_{\lambda_j} < 2 A^*_j$.
\end{myproofname}

\begin{myproofname}{Theorem}{\ref{prop:delta_inequality}}
\begin{itemize}
\item[(i)]
 For any $k, l \in N$, if $r_{o(k)} \ge r_{o(l)} $, then we have that
 \begin{alignat*}{2}
	\frac{a_{o(l)} ( p_{o(k)}-a_{o(k)})} { p_{o(l)} p_{o(k)} }
	& = r_{o(l)}(1- r_{o(k)}) \\
	& \le r_{o(k)} ( 1 - r_{o(l)} ) \\
	& = \frac{a_{o(k)} ( p_{o(l)}-a_{o(l)})} { p_{o(l)} p_{o(k)} }.
	\end{alignat*}
%
%
\item[(ii)]
Suppose that $k< l$ and $ r_{o(k)} < r_{o(l)} $ for $k, l \in N$.
If $a_{o(l)} > A^{\xo}_{l+1}$, then we have from Definition \ref{def.o(a,p)}-(ii) that
\[
  a_{o(k)} \ge a_{o(l)} > A^{\xo}_{l+1}.
\]
Otherwise, $a_{o(l)} \le A^{\xo}_{l+1}$. If $a_{o(k)} \le A^{\xo}_{l+1}$,
we see that $ r_{o(k)} \ge r_{o(l)} $ from Definition \ref{def.o(a,p)}-(iii), which contradicts to the assumption $ r_{o(k)} < r_{o(l)} $.
Hence we have that
\[
  a_{o(k)} > A^{\xo}_{l+1} \ge a_{o(l)}.
\]
\end{itemize}
\end{myproofname}

\section{Proof of Theorems \ref{thm:just_in_time} and \ref{thm:translation}}
%
Note that Theorems \ref{thm:just_in_time} and \ref{thm:translation} are proved in \cite{hashimoto2021tight}
for a simple list scheduling algorithm when $p_i=1 (i \in N)$ and the approximation ratio is 2.
Although the proofs of the theorems are similar to those in \cite{hashimoto2021tight},
we prove them for confirmation.
The next three lemmas are used in the proof of Theorem \ref{thm:just_in_time}.
\begin{lem} (Lemma 1 in \cite{hashimoto2021tight})
	\label{lem:add1}
Let $\xa=(a_1,a_2,\dots,a_n)^\trans$, $\xv=(v_1,v_2,\dots,v_n)^\trans$,
$\xb=(b_1,b_2,\dots,b_q)^\trans$, and $\xu=(u_1,u_2,\dots,u_q)^\trans$ be nonnegative vectors such that $\sum_{i=1}^n a_i = \sum_{i=1}^q b_i$.
If
\[
  \sum_{i:v_i \le t} a_i \le \sum_{i:u_i \le t} b_i \quad \mbox{for any} \quad t \ge 0,
\]
then
$
  \xb^\trans \xu \le \xa^\trans \xv.
$
\end{lem}

The next two lemmas generalize results in \cite{hashimoto2021tight}, where $p_i=1 (i \in N)$. 
\begin{lem}
	\label{lem:optlemma}
Let $ S^\prime = \langle n, \mj, \xp, \xa,n, (\xC - \xp), \xa\rangle $ be any instance of NR-SSP,
where $ \xC=(C_1, C_2, \dots, C_n)^\trans $ 
such that $ C_1 \ge p_1 $
and $ C_{j+1}-C_j\ge p_{j+1} $ for any $ j \in \{1,2,\dots, n-1\} $,
then $\xC$ is optimal for $ S^\prime $.
\end{lem}
\begin{myproof}
It is clear that $ \xC $ is a feasible schedule for $ S^\prime $.
Let $\xC^v$ be any feasible schedule for $ S^\prime $. Then we have that
\[
 \sum_{j:C^v_j - p_j \le T}a_j \le \sum_{i:u_i \le T}b_i \ \mbox{for any} \ T \ge 0
\]
from \eqref{eq:balance}, where $\xu=\xC - \xp$ and $\xb=\xa$.
We obtain
\[
 \xa^\trans (\xC- \xp) \le \xa^\trans( \xC^v- \xp),
\]
by applying Lemma \ref{lem:add1} for $\xa$, $\xv=\xC^v - \xp$, $\xb=\xa$, and $\xu=\xC- \xp$.
Since the object function value is $\xa^\trans \xC^v$, $\xC$ is optimal for $ S^\prime $.
\end{myproof}

\begin{lem}
	\label{lem:greedyineq}
	Let $ S = \langle n,\mj, \xp, \xa, q, \xu, \xb\rangle$ and $ \tilde{S} = \langle n, \mj, \xp, \xa,q^\prime, \xu^\prime, \xb^\prime\rangle$ be two instances of NR-SSP.If
\begin{equation} \label{eq:add20}
 {\displaystyle \sum_{i:u^\prime_i \le T}b^\prime_i \le \sum_{i:u_i \le T}b_i} \ \mbox{for any} \ T\ge0,
\end{equation}
then $\xC^{\xo}(S) \le \xC^{\xo}(\tilde{S}) $ and
$f_S(\xC^{\xo}(S)) \le f_{\tilde S}(\xC^{\xo}(\tilde{S}))$ for any $ \xo \in {\mathcal O} $.
\end{lem}
\begin{myproof}
From condition \eqref{eq:add20}, Algorithm \ref{alg:list} for any permutation $ \xo $ outputs $\xC^{\xo}(S)$ for $S$ and $\xC^{\xo}(\tilde{S})$ for $\tilde{S}$ so that
\[
 \xC^{\xo}(S) \le \xC^{\xo}(\tilde{S}).
\]
Then we easily see that $f_S(\xC^{\xo}(S)) \le f_{\tilde S}(\xC^{\xo}(\tilde{S}))$.
\end{myproof}

\begin{myproofname}{Theorem}{\ref{thm:just_in_time}}
Recall that $S=\langle n,\mj, \xp, \xa, q, \xu, \xb \rangle$ and $ S^\prime = \langle n,\mj, \xp, \xa, n, (\xC^*(S) - \xp), \xa\rangle$ in Theorem \ref{thm:just_in_time}.
From Lemma \ref{lem:optlemma},
$ \xC^*(S) $ is an optimal schedule for $ S^\prime$.
Since $ \xC^*(S) $ is a feasible schedule for $ S $, \eqref{eq:balance} holds for every $ T \ge 0 $,
namely,
\[
  \sum_{j:C^*(S)_j-p_j \le T}a_j \le \sum_{i:u_i \le T}b_i  \ \mbox{for any} \  T\ge 0.
\]
Hence $ f_{S}(\xC^{\xo}(S)) \le f_{S^\prime}(\xC^{\xo}(S^\prime))$ holds for any $ \xo\in {\mathcal O} $ from Lemma \ref{lem:greedyineq}.
If $f_{S^\prime}(\xC^{\xo}(S^\prime)) < 3 f_{S^\prime}(\xC^*(S^\prime))$, then we see that
	\begin{alignat*}{2}
	 f_S(\xC^{\xo}(S)) & \le f_{S^\prime}(\xC^{\xo}(S^\prime))\\
	 & < 3 f_{S^\prime}(\xC^*(S^\prime))\\
	 & = 3 {\bm a^\trans} \xC^*(S)\\
	 & = 3 f_S(\xC^*(S)).
	\end{alignat*}
\end{myproofname}

Then we prove Theorem \ref{thm:translation} by using the following lemma.
\begin{lem}
	\label{lem:suff3opt}
Let $ S = \langle n,\mj, \xp, \xa, n, (\xC - \xp), \xa\rangle$ be any instance of NR-SSP,
where $ \xC =(C_1,C_2,\dots,C_n)^\trans $ such that $ C_1 \ge p_1 $ and $ C_{j+1}-C_j\ge p_{j+1} $ for any $ j \in \{1,2,\dots, n-1 \}$.
Let $ \xd_k = (\xzero_k^\trans ~\xone_{n-k}^\trans)^\trans$ be a vector whose first $k$ elements are $0$ and the latter $n-k$ elements are $1$ for an integer $ k \in \{0,1,\dots,n-1\}$.
Define $ S^\dagger = \langle n, \mj, \xp, \xa, n, (\xC + L\xd_k - \xp ), \xa\rangle $ for any $L \ge 0$.

For any $ \xo\in O(\xa, \xp) $, $ f_{S}(\xC^{\xo}(S)) < 3f_{S}(\xC^{*}(S))$ implies $ f_{S^\dagger}(\xC^{\xo}(S^\dagger)) < 3f_{S^\dagger}(\xC^{*}(S^\dagger)) $.
\end{lem}
\begin{myproof}
Clearly $\xC$ is a feasible schedule for $ S $.
	By Lemma \ref{lem:optlemma}, we have $ \xC^*(S)=\xC $ and $\xC^*(S^\dagger)=\xC + L\xd_k$.
Then we see
\begin{alignat*}{2}
	f_{S^\dagger}(\xC^*(S^\dagger)) - f_{S}(\xC^*(S)) &= \xa^\trans (L \xd_k)\\
	&= L\sum_{j=k+1}^na_j\\
	&= LA^*_{k+1}.
\end{alignat*} 
Since supply times $C_j-p_j, j \in \{1,2,\dots,k\}$ in $ S^\dagger$ are equal to those in $S$,
we see from Algorithm \ref{alg:list} that
\[ 
	C^{\xo}_{o(j)}(S) = C^{\xo}_{o(j)}(S^\dagger) \quad \mbox{for any} \quad j \in \{1,2,\dots, \alpha_k \},
\] 
where
\begin{alignat*}{2}
	\alpha_{k} & = \max \{ j \in N | a_{o(1)} + \cdots + a_{o(j)} \le a_1 + \cdots + a_{k} \} \\
	& = \min \{ j \in N | a_{o(j+1)} + \cdots + a_{o(n)} \ge a_{k+1} + \cdots + a_{n} \} \\
	& = \lambda_{k+1} -1.
\end{alignat*}
Then $ \xC^\prime =(C^\prime_1, C^\prime_2,\dots,C^\prime_n)$ define by
\[
  C^\prime_{o(j)} = \left\{
  \begin{array}{ll}
  C^{\xo}_{o(j)}(S) & \quad \mbox{for any} \quad j \in \{1,2,\dots, \lambda_{k+1}-1\} \\
  C^{\xo}_{o(j)}(S) +L & \quad \mbox{for any} \quad j \in \{\lambda_{k+1},\dots, {n} \}\
  \end{array}
  \right.
\]
is a feasible schedule for $ S^\dagger $ since $\xC^{\xo}(S)$ is feasible for $ S $.
%
We easily see that $\xC^{\xo}(S^\dagger) \le \xC^\prime$
and $f_{S^\dagger}(\xC^{\xo}(S^\dagger) )\le f_{S^\dagger}(\xC^\prime)$ from Algorithm \ref{alg:list}.
Then we have that
\begin{alignat*}{2}
	 f_{S^\dagger}(\xC^{\xo}(S^\dagger)) - f_{S}(\xC^{\xo}(S))
  &\le f_{S^\dagger}(\xC^\prime) - f_{S}(\xC^{\xo}(S))\nonumber\\
	& = \xa^\trans (\xC^\prime - \xC^{\xo}(S))\nonumber\\
	& = L\sum_{j=\lambda_{k+1}}^n a_{o(j)}\nonumber \\
	& = LA^{\xo}_{\lambda_{k+1}}\nonumber \\
	& < 2LA^*_{k+1}\\
	& = 2 (f_{S^\dagger}(\xC^*(S^\dagger)) - f_{S}(\xC^*(S))),
\end{alignat*}
where the strict inequality follows from Theorem \ref{prop:2opt_condition}.
Therefore if $ f_{S}(\xC^{\xo}(S)) < 3f_{S}(\xC^{*}(S)) $, then we have that
	 \begin{alignat*}{2}
  &f_{S^\dagger}(\xC^{\xo}(S^\dagger))\\
 &< f_{S}(\xC^{\xo}(S)) + 2 (f_{S^\dagger}(\xC^*(S^\dagger)) - f_{S}(\xC^*(S))) \\
  & < 3f_{S^\dagger}(\xC^{*}(S^\dagger)).
	 \end{alignat*}
\end{myproof}

\begin{myproofname}{Theorem}{\ref{thm:translation}}
Recall that $ S = \langle n, \mj, \xp, \xa,n, (\xC^\prime - \xp), \xa\rangle $ and $ \bar{S} = \langle n, \mj, \xp, \xa, n, (\xsigma(\xp) - \xp), \xa\rangle$ in Theorem \ref{thm:translation}.
We can express that
\begin{alignat*}{2}
  \xC^\prime - \xsigma(\xp) = & (C^\prime_1 -C^\prime_0 -p_1) \xd_0 + (C^\prime_2- C^\prime_1 - p_2) \xd_1 \\
  & + \cdots + (C^\prime_n- C^\prime_{n-1} - p_n) \xd_{n-1} \\
  = & \sum_{k=0}^{n-1} L_k \xd_k,
\end{alignat*}
where $C^\prime_0=0$, $ \xd_k = (\xzero_k^\trans ~\xone_{n-k}^\trans)^\trans $,
and $L_k=C^\prime_{k+1}- C^\prime_k - p_{k+1}$ for any $k \in \{0,1,\ldots,n-1\}$.
From the conditions in the theorem, we have that $L_k \ge 0$ for any $k \in \{0,1,\ldots,n-1\}$.
We define $\xD_0=\xsigma(\xp)$ and
\[
 \xD_k = \xD_{k-1}+ L_{k-1} \xd_{k-1}
\]
for any $k \in N$.
Then we see that
\[
  \xD_{n} = \xC^\prime.
\]
We also define $(n+1)$-instances $S^k=\langle n, \mj, \xp, \xa,n, (\xD_k - \xp), \xa\rangle$
for $k \in \{0,1,\ldots,n\}$.
Then we have that $S^0= \bar{S}$ and $S^n=S$.

Now we consider two instances $S^0 = \bar{S}$ and $S^1$.
If $ f_{S^0}(\xC^{\xo}(S^0)) < 3f_{S^0}(\xC^{*}(S^0)) $, we have $ f_{S^1}(\xC^{\xo}(S^1)) < 3f_{S^1}(\xC^{*}(S^1)) $
from Lemma \ref{lem:suff3opt}.
For $k=2,\ldots,n$, we repeat this procedure for two instances $S^{k-1}$ and $S^k$, then we
finally obtain that $ f_{S^n}(\xC^{\xo}(S^n)) < 3f_{S^n}(\xC^{*}(S^n))$, where $S^n=S$.
\end{myproofname}

\section{Proof of Theorems \ref{thm:main}, \ref{thm:lowerbound}, and \ref{thm:same_ratio} }

In this section, we prove Theorem \ref{thm:same_ratio} at first.
Then Theorem \ref{thm:main} is proved by using Theorems \ref{thm:just_in_time}, \ref{thm:translation}, and \ref{thm:same_ratio}.
At last, we prove Theorem \ref{thm:lowerbound} by showing a simple instance.

\begin{myproofname}{Theorem}{\ref{thm:same_ratio}}
Recall that $S = \langle n, \mj, \xp, \xa, n, (\xsigma(\xp) - \xp), \xa\rangle$ and $ S^\dagger = \langle n, \mj, \xa, \xa, n, (\xsigma(\xa) - \xa), \xa\rangle$.
We define ${\bm \delta} = (\delta_1, \delta_2, \dots, \delta_n)^\trans$ by
\[
  \delta_j = p_j - a_j \ \mbox{for each} \ j \in N.
\]
From Assumption \ref{asum:min_ratio}, $\delta_{j} \ge 0$ for any $j \in N$.
Then we easily see that $O(\xa,\xp) \subset O(\xa,\xa)$.
For any $\xo \in O(\xa, \xp) $, we need to prove
\begin{equation} \label{eq:theorem5-1}
	f_{S^\dagger}(\xC^{\xo}(S^\dagger)) < 3f_{S^\dagger}(\xC^{*}(S^\dagger)) \Rightarrow f_{S}(\xC^{\xo}(S)) < 3f_{S}(\xC^{*}(S)).
\end{equation}
%
Define $ D^*(S) = f_{S}(\xC^*(S)) - f_{S^\dagger}(\xC^*(S^\dagger))$ and $ D^{\xo}(S) = f_{S}(\xC^{\xo}(S)) - f_{S^\dagger}(\xC^{\xo}(S^\dagger))$.
To prove (\ref{eq:theorem5-1}), it suffices to show that
\begin{equation}
	\label{eq:diff_optimal_mu}
	D^{\xo}(S) \le 3D^*(S).
\end{equation}
Since the weight vector $ \xw=\xa $ is common both in $S$ and $S^\dagger$, we have that
\begin{alignat*}{2}
	D^*(S) &= \xa^\trans(\xC^*(S) - \xC^*(S^\dagger)),\\
	D^{\xo}(S) &= \xa^\trans(\xC^{\xo}(S) - \xC^{\xo}(S^\dagger)).
\end{alignat*}
From Lemma \ref{lem:optlemma}, $\xC^*(S) = \xsigma(\xp) $ and $ \xC^*(S^\dagger) = \xsigma(\xa) $.
Then we see that
\begin{alignat}{2}
	\label{eq:dopt}
	D^*(S) &= \xa^\trans(\xsigma(\xp) - \xsigma(\xa)) \nonumber\\
	& = \sum_{j=1}^n a_j \sum_{k= 1}^j \delta_k \nonumber\\
	&= \sum_{k=1}^{n} \delta_{k} \sum_{j=k}^{n} a_{j}\nonumber\\
	&= \sum_{k=1}^{n}\delta_{o(k)} \sum_{j=o(k)}^{n} a_{j} \nonumber\\
	&=\sum_{k=1}^{n} A^*_{o(k)} \delta_{o(k)}.
\end{alignat}

Next we evaluate $ D^{\xo}(S) $. Comparing $ S^\dagger $ with $ S$, the processing time vector and the supply time vector are different.
Let $k \in N$.
When $a_{k}$ increases by $\delta_{k}$, $j$-th element of the supply time vector $\xu=\xsigma(\xa)-\xa$ increases by $\delta_{k}$ for each $j \in \{k+1,\dots,n\}$.
Then $C^{\xo}_{o(1)}, \dots, C^{\xo}_{o(\alpha_{k})} $ do not change and
$C^{\xo}_{o(\alpha_{k}+1)}, \dots, C^{\xo}_{o(n)} $ may increase at most $ \delta_{k} $, where
\begin{alignat*}{2}
	\alpha_{k} &= \max \{ j \in N | a_{o(1)} + \cdots + a_{o(j)} \le a_1 + \cdots + a_{k} \}\\
	 &= \lambda_{k+1} -1.
\end{alignat*}
Figure \ref{fig:supply} illustrates this situation.
\begin{figure*}[t]
	\centering
	\includegraphics[width=0.9\textwidth]{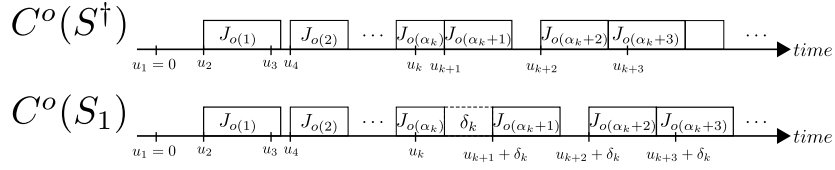}
	\caption{Gantt-charts of the schedules by the algorithm for $ S^\dagger $ and $ S_1 $, where the supply times $u_j$ ($j \ge k+1$) in $S_1$ increase from those in $S^\dagger$ by $\delta_k$.}
	\label{fig:supply}
\end{figure*}

Besides, when $ o(k) $-th element of processing time vector increases by $ \delta_{o(k)} $, not only $ C^{\xo}_{o(k)}$ but also $C^{\xo}_{o(k+1)}, \dots, C^{\xo}_{o(n)} $ may increase at most $ \delta_{o(k)} $.
Figure \ref{fig:processing} depicts this situation.
\begin{figure*}[t]
	\centering
	\includegraphics[width=0.9\textwidth]{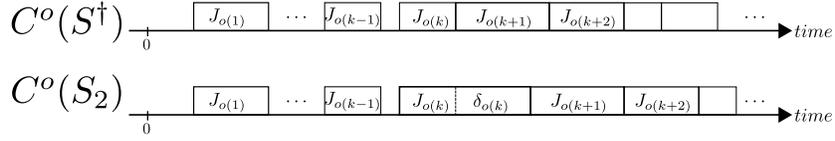}
	\caption{Gantt-charts of the schedules by the algorithm for $ S^\dagger $ and $ S_2 $, where the processing time $p_{o(k)}$ in $S_2$ increases from that in $S^\dagger$ by $\delta_o(k)$.}
	\label{fig:processing}
\end{figure*}
From these facts, we have that
\begin{alignat}{2}
	D^{\xo}(S) &\le \sum_{k=1}^{n} \delta_{k} \sum_{l=\alpha_{k} + 1}^n a_{o(l)} + \sum_{k=1}^{n} \delta_{o(k)}\sum_{l=k}^na_{o(l)} \nonumber\\
	&= \sum_{k=1}^{n} A^{\xo}_{\alpha_{k} + 1} \delta_{k} + \sum_{k=1}^{n} A^{\xo}_k \delta_{o(k)} \nonumber\\
	&= \sum_{k=1}^{n} A^{\xo}_{\lambda_{k + 1}} \delta_{k} + \sum_{k=1}^{n} A^{\xo}_k \delta_{o(k)} \nonumber\\
	&= \sum_{k=1}^{n} A^{\xo}_{\lambda_{o(k) + 1}} \delta_{o(k)} + \sum_{k=1}^{n} A^{\xo}_k \delta_{o(k)} \nonumber\\
	&\le \sum_{k=1}^{n} \left(2A^{*}_{o(k)+1} + A^{\xo}_{k}\right)\delta_{o(k)},  \label{eq:do1}
\end{alignat}
where the last inequality follows from Theorem \ref{prop:2opt_condition}.
Comparing \eqref{eq:do1} with \eqref{eq:dopt}, the inequality
\[ 
	\sum_{k=1}^{n}A^{\xo}_{k}\delta_{o(k)} \le \sum_{k=1}^{n} \left(2a_{o(k)} + A^*_{o(k)}\right)\delta_{o(k)},
\] 
implies \eqref{eq:diff_optimal_mu}. The inequlity above is proved in Lemma \ref{lem:proof-th1} below.
\end{myproofname}

\begin{lem} \label{lem:proof-th1}
For any instance $S = \langle n, \mj, \xp, \xa, n, (\xsigma(\xp) - \xp), \xa\rangle$ and $\xo \in O(\xp,\xa)$, we define
\begin{alignat*}{2}
&M^*(S)= \sum_{k=1}^{n} \left(2a_{o(k)} + A^*_{o(k)}\right)\delta_{o(k)}\\
&\mbox{and}\\
&M^{\xo}(S)=\sum_{k=1}^{n}A^{\xo}_{k}\delta_{o(k)},
\end{alignat*}
where $\delta_j = p_j - a_j$ for each $j \in N$.
Then we have that
\begin{equation}
	M^{\xo}(S) \le M^*(S). \label{eq:diff_m}
\end{equation}
\end{lem}
\begin{myproof}
We show \eqref{eq:diff_m} by induction on $ n $. In the case $ n=1 $, the result is trivial.
Then we prove \eqref{eq:diff_m} for $ S $, $ \xo $, and $ n \ge 2 $ under the assumption that \eqref{eq:diff_m} holds for $n-1$.

We define $ S^- = \langle n-1, \mj^-, \xp^-, \xa^-, n-1, (\xsigma(\xp^-) - \xp^-), \xa^-\rangle$ such that $\mj^- = (J_1, J_2, \dots, J_{o(1)-1}, J_{o(1)+1}, \dots, J_n)$, $\xp^- = (p_1, p_2, \dots, p_{o(1)-1}, p_{o(1)+1}, \dots, p_n)^\trans$, and $\xa^- = (a_1, a_2, \dots, a_{o(1)-1}, a_{o(1)+1}, \dots, a_n)^\trans$.
We also define $ \xo^- = (o(2), o(3), \dots, o(n)) $, $ \Delta^* = M^*(S) - M^*(S^-) $, and $ \Delta^{\xo} = M^{\xo}(S) - M^{\xo^-}(S^-) $.
It is obvious that $ \xo^-$ satisfies the conditions in Definition \ref{def.o(a,p)}.
Under the assumption $ M^{\xo^-}(S^-) \le M^*(S^-) $, we need to show
\begin{equation}
	\label{eq:Delta}
	\Delta^{\xo} \le \Delta^{*}
\end{equation}
to prove \eqref{eq:diff_m} for $n$.

Next we show that \eqref{eq:Delta} holds.
We have that
\begin{alignat*}{2}
	 M^*(S^-)
	 & = \sum_{k=2}^{n} \left(2a_{o(k)} + \sum_{o(k) \le j \le n, j \ne o(1)} a_j \right)\delta_{o(k)} \\
\end{alignat*}
and
\begin{alignat*}{2}
	\Delta^{*} &= M^*(S) - M^*(S^-)\nonumber\\
	&= 2a_{o(1)}\delta_{o(1)} + A^*_{o(1)}\delta_{o(1)} + \sum_{o(k) < o(1)}a_{o(1)}\delta_{o(k)}. 
\end{alignat*}
We also have that
\begin{alignat*}{2}
	 M^{\xo^-}(S^-)
	 & = \sum_{k=2}^{n} \delta_{o(k)} \sum_{j=k}^n a_{o(j)}
\end{alignat*}
and
\begin{alignat*}{2}
	\Delta^{\xo} &= M^{\xo}(S) - M^{\xo^-}(S^-)\nonumber\\
	&= \sum_{j=1}^n a_{o(j)} \delta_{o(1)} \nonumber \\
	& = \sum_{j=1}^{j^*-1} a_{o(j)} \delta_{o(1)} + a_{o(j^*)} \delta_{o(1)} + A^{\xo}_{j^*+1} \delta_{o(1)}, 
\end{alignat*}
where
\[
  j^* = \min \{ j \in N | r_{o(j)} > r_{o(1)} \}.
\]
(In the case of $r_{o(j)} \le r_{o(1)}$ for any $j \in N$, we define $j^*=n+1$ and $a_{o(j^*)}=0$.)
Since $r_{o(j)} \le r_{o(1)}$ for any $j < j^*$, we have from Theorem \ref{prop:delta_inequality}-(i) that
\begin{alignat*}{2}
 \sum_{j=1}^{j^*-1} a_{o(j)} \delta_{o(1)} 
	 &=\sum_{1 \le j \le j^*-1, o(j) \ge o(1) } a_{o(j)} \delta_{o(1)}
	 + \sum_{1 \le j \le j^*-1, o(j) < o(1)} a_{o(j)} \delta_{o(1)} \\
	&\le\sum_{1 \le j \le j^*-1, o(j) \ge o(1) } a_{o(j)} \delta_{o(1)}
	+ \sum_{1 \le j \le j^*-1, o(j) < o(1)} a_{o(1)} \delta_{o(j)} \\
	&\le A^*_{o(1)}\delta_{o(1)} + \sum_{o(k) < o(1)}a_{o(1)}\delta_{o(k)}.
\end{alignat*}
Since $j^* > 1$ and $r_{o(j^*)} > r_{o(1)}$, we have from Theorem \ref{prop:delta_inequality}-(ii) that
\begin{alignat*}{2}
  a_{o(j^*)} \delta_{o(1)} + A^{\xo}_{j^*+1} \delta_{o(1)}
  & \le 2 a_{o(1)} \delta_{o(1)}.
\end{alignat*}
Hence $\Delta^{\xo} \le \Delta^*$ holds.
\end{myproof}

\begin{myproofname}{Theorem}{\ref{thm:main}}
By Theorems \ref{thm:just_in_time}-\ref{thm:same_ratio}, it suffices to prove it for any instance
$ S = \langle n, \mj, \xa, \xa, n, (\xsigma(\xa) - \xa), \xa\rangle $ and $\xo \in O(\xa,\xp)$.
From Lemma \ref{lem:optlemma}, the optimal schedule for $S$ is $\xC^*(S) = \xsigma(\xa) $. Define $ Q = \sum_{j=1}^n a_j $.
Then we see that
\begin{alignat*}{2}
	f_S(\xC^*(S)) &= \xa^\trans \xsigma(\xa) \nonumber \\
	& = \sum_{j=1}^n a_j \sum_{i=1}^j a_i  \nonumber \\
	& ={1 \over 2} Q^2 + {1 \over 2}\sum_{j=1}^n a_j^2.
\end{alignat*}

The output schedule of Algorithm 2 is $\xC^{\xo}(S)$ for $\xo \in O(\xa, \xp)$.
Since all the resources are available after the time $t=Q - a_{o(n)} $,
we have that
\[
  C_{o(j)}^{\xo}(S) \le Q - a_{o(n)} + \sum_{i=1}^j a_{o(i)} < Q + \sum_{i=1}^j a_{o(i)}
\]
for each $j \in N$. Thus we see that
\begin{alignat}{2}
	f_S(\xC^{\xo}(S))
	&= \sum_{j=1}^n a_{o(j)} C_{o(j)}^{\xo}(S) \nonumber \\
	& < \sum_{j=1}^n a_{o(j)} \left( Q + \sum_{i=1}^j a_{o(i)} \right) \nonumber \\
	& = Q^2 + \sum_{j=1}^n a_{o(j)} \sum_{i=1}^j a_{o(i)} \nonumber \\
	& = Q^2 + {1 \over 2} Q^2 + {1 \over 2}\sum_{j=1}^n a_j^2 \nonumber \\
	&< 3f_S(\xC^*(S)) \label{eq:thm1_3}. \nonumber
\end{alignat}
\end{myproofname}

Finally we prove Theorem \ref{thm:lowerbound}.
\begin{myproofname}{Theorem}{\ref{thm:lowerbound}}
Let $ 0< e< 0.1 $ and $ \tilde{S} $ = $ \langle n, \mj, \xp, \xa, n, (\xsigma(\xp)-\xp), \xa \rangle $ for $n=3$, $ \xp = (e, e, 1)^\trans $, and $ \xa=(1-e, 1, 1+e)^\trans $.
We easily see that
\[
 \begin{array}{l}
  \xC^*(\tilde{S}) = \xsigma(\xp) = (e,2e,1+2e)^\trans, \\
  \xo = (3,2,1), \\
 \xC^{\xo}(\tilde{S}) = (1+3e,1+2e,1+e)^\trans.
 \end{array}
\]
Then
\begin{alignat*}{2}
 &\quad\lim\limits_{e\to 0^+}\frac{f_{\tilde{S}}(\xC^{\xo}(\tilde{S}))}{f_{\tilde{S}}(\xC^*(\tilde{S}))}\\
 &= \lim\limits_{e\to 0^+}\frac{(1-e)(1+3e) + (1+2e) + (1+e)(1+e)}{(1-e)e + 2e + (1+e)(1+2e)}\\
 &=3.
\end{alignat*}
Hence we obtain the result.

\end{myproofname}

\bibliography{cite}
\end{document}